\documentclass[a4paper,11pt]{article}
\usepackage{pos}
\usepackage{wrapfig}

\title{
    Renormalization group on tensor networks
}
\ShortTitle{RG on tensor networks}

\usepackage{tikz,xcolor}
\definecolor{lime}{HTML}{A6CE39}
\DeclareRobustCommand{\orcidicon}{%
	\begin{tikzpicture}
	\draw[lime, fill=lime] (0,0) 
	circle [radius=0.16] 
	node[white] {{\fontfamily{qag}\selectfont \tiny ID}};	\draw[white, fill=white] (-0.0625,0.095) 
	circle [radius=0.007];	\end{tikzpicture}
	\hspace{-2mm}}
\foreach \x in {A,B,C,D,E}{%
	\expandafter\xdef\csname orcid\x\endcsname{\noexpand\href{https://orcid.org/\csname orcidauthor\x\endcsname}{\noexpand\orcidicon}}
}

\newcommand{\TU}{\affiliation[a]{Center for Computational Sciences, University of Tsukuba, Tsukuba, Ibaraki 305-8577, Japan}}

\newcommand{\UTokyo}{\affiliation[b]{Graduate School of Science, The University of Tokyo, Bunkyo-ku, Tokyo, 113-0033, Japan}}

\author*[a,b]{Shinichiro Akiyama~\orcidA{}}
\emailAdd{akiyama@ccs.tsukuba.ac.jp}
\TU
\UTokyo

\notes{
    \note{UTHEP-820, UTCCS-P-177}
}

\abstract{
We review recent developments in tensor network approaches, focusing on renormalization group methods.
Since they are free from the negative sign and complex action problems, there is growing interest in their application to lattice field theories, particularly with a view toward future studies of quantum chromodynamics (QCD) at finite temperature and density.
They are also of broad interest in quantum field theory, with recent advances in approaches that allow one to directly investigate universal aspects of critical behavior by making use of theoretical insights from conformal field theory.
We highlight several recently explored topics that are expected to play important roles in forthcoming tensor-network studies of QCD.
}

\FullConference{The 42nd International Symposium on Lattice Field Theory (LATTICE2025)\\
2-8 November 2025\\
Tata Institute of Fundamental Research, Mumbai, India\\}

\begin{document}
\maketitle

\section{Introduction}
\label{sec:Intro}

Tensor networks provide a powerful alternative numerical framework for studying lattice field theories.
Originally developed in statistical and condensed-matter physics, tensor networks have recently attracted increasing attention in high-energy physics as well.
A notable advantage of these approaches is that they are free from the infamous sign problem. 
For this reason, their application to quantum chromodynamics (QCD) at finite temperature and density is highly anticipated.
Moreover, tensor networks serve as a natural bridge between classical and quantum computations, since quantum circuits can be straightforwardly mapped onto tensor-network representations, which can be handled on classical computers.
In this context, tensor networks are also expected to provide a powerful framework for quantum-classical hybrid algorithms, thereby advancing the study of real-time dynamics in quantum many-body systems.

In this review, we focus on tensor networks from the perspective of a practical formulation of the renormalization group (RG), specifically as a kind of real-space RG.
Within this framework, the partition function of a quantum many-body system can be evaluated by extracting the information relevant to low-energy physics from its tensor-network representation.
This approach provides a valuable alternative for systems to which conventional Monte Carlo (MC) simulations do not apply.

Historically, the block-spin transformation introduced by Kadanoff~\cite{Kadanoff:1966wm} opened an avenue for real-space RG as a practical method for understanding many-body systems.
Building on the idea of scale transformations, the theory of the RG was established.
The basic concept of the RG was clearly articulated by Wilson and Kogut~\cite{Wilson:1973jj}: 
The essential idea is the iterative application of a transformation $\tau$ acting on an effective Hamiltonian $\mathcal{H}^{(l)}$ at scale $l$, such that $\tau(\mathcal{H}^{(l)}) = \mathcal{H}^{(l+1)}$. 
Each application of $\tau$ reduces the degrees of freedom by a factor of $2$ while simultaneously doubling the length scale.
The transformation is repeated $l$ times until the coarse-grained length scale, increased by a factor of $2^l$, becomes comparable to the correlation length.
Under repeated application, $\mathcal{H}^{(l)}$ approaches a fixed point.
As we will see, RG methods based on tensor networks reformulate this iterative transformation in the language of tensors, rather than in terms of Hamiltonians.

Wilson also pioneered the use of the RG by applying it to the single-impurity Kondo problem~\cite{Wilson:1974mb}, an effectively 0D problem.
The extension to 1D quantum systems was subsequently achieved with the density matrix RG (DMRG)~\cite{White:1992zz}.
The DMRG paradigm is a systematic reduction of the number of relevant states within the density-matrix formalism. 
DMRG can be understood as a variational method formulated in terms of matrix product states (MPS), which provide an efficient representation for gapped systems~\cite{Schollwoeck:2010uqf}, reflecting the area law of entanglement entropy~\cite{Eisert:2008ur}. 
MPS also form a broad class of tensor-network algorithms.
Their applications to lattice gauge theories (LGT) have accelerated~\cite{Byrnes:2002nv,Banuls:2013jaa,Itou:2024psm,Fujii:2024reh,ArguelloCruz:2024xzi}, driven by the growing importance of Hamiltonian-based simulations in the wake of recent progress in quantum computing.

The transfer-matrix formalism also allows for a similar approach within the Lagrangian framework.
A prominent example is the corner transfer matrix RG (CTMRG)~\cite{baxter1978variational,Nishino_1996}, which provides an efficient way for evaluating partition functions of 2D classical spin systems.
\footnote{
Recently, CTMRG has been applied to investigate the phase structure of the Gross--Neveu model with Wilson fermions within the Lagrangian formalism~\cite{Kong:2026ofl}. 
CTMRG is also widely used in Hamiltonian-based simulations as an approximate contraction method for the 2D tensor networks that arise in the evaluation of norms and expectation values described by tensor network states.
}
CTMRG can be regarded as a density-matrix-based variational approach grounded in the quantum-classical correspondence.
Building further on this line of ideas, Levin and Nave introduced a variant of real-space RG for 2D classical systems, motivated by the broader objective of generalizing RG approaches to higher dimensions~\cite{Levin:2006jai}.
The approach they introduced constitutes a systematically improvable real-space RG.
Such methods are now collectively referred to as the tensor RG (TRG).
Interestingly, the TRG can be regarded as a concrete realization of the RG concept articulated above by Wilson and Kogut.
\footnote{
Precisely speaking, the original Levin–Nave TRG does not necessarily converge to a physical fixed point, since short-range correlations in the partition function are not properly removed during the RG transformation.
By incorporating additional procedures to eliminate these correlations, a genuine RG fixed point can be obtained numerically~\cite{Evenbly:2015ucs,Yang:2017lvo}.
TRG schemes augmented in this way are collectively referred to as tensor network renormalization (TNR).
We note, however, that the terminology is not always consistent.
In some literature, both TRG and TNR are grouped together under the name tensor-network RG (TNRG) methods.
}
The first application of TRG to LGT was initiated in 2014 by Shimizu and Kuramashi in a series of works investigating the Schwinger model~\cite{Shimizu:2014uva,Shimizu:2014fsa,Shimizu:2017onf}.
As envisioned by Levin and Nave, several algorithms suitable for higher-dimensional systems have since been proposed~\cite{Xie:2012mjn,Adachi:2019paf,Kadoh:2019kqk,Ueda:2025mhu}, and TRG has been applied to both (2+1)D~\cite{Unmuth-Yockey:2018xak,Kuramashi:2018mmi,Bloch:2021mjw,Bloch:2021uup,Kuwahara:2022ubg,Akiyama:2024qgv,Yosprakob:2024sfd,Luo:2025qtv,Naravane:2026zjk} and (3+1)D theories~\cite{Akiyama:2019xzy,Akiyama:2020ntf,Akiyama:2020soe,Milde:2021vln,Akiyama:2021zhf,Akiyama:2022eip,Akiyama:2023hvt,Sugimoto:2025vui,Sugimoto:2026wnw}.
The community has also achieved a number of developments that are of intrinsic interest from a broader quantum field theory perspective, including connections to conformal field theory (CFT) and critical phenomena.

This review is organized as follows. 
We first provide a brief overview of the basic formulation of TRG in Sec.~\ref{sec:formulation}.
The subsequent Sections~\ref{sec:GTN}--\ref{sec:synergy} review selected recent theoretical and numerical advances along the directions described above.
Sec.~\ref{sec:summary} is devoted to a summary.

\section{Basic formulation of TRG}
\label{sec:formulation}

The TRG approach consists of two main ingredients.
First, the partition function $Z$ is expressed as a network of tensors.
This tensor network is defined through the contraction of local tensors, typically associated with each lattice site.
The tensor-network representation of $Z$ usually inherits the same geometry as the lattice on which the theory is defined.
Suppose that an ultra-local theory is defined on a $d$-dimensional lattice $\Lambda_{d}$ with periodic boundary conditions.
Under these assumptions, the partition function $Z$ of the theory can be represented in the form of $Z={\rm tTr}\left[\prod_{n\in\Lambda_{d}}T_{n}\right]$, where $T_{n}$ denotes a fundamental tensor located at site $n$, and its indices are omitted for notational simplicity.
Typically, tensors $T_{n}$ can be defined such that their rank matches the coordination number of $\Lambda_d$.
The notation ``${\rm tTr}$" indicates that all tensors $T_{n}$ are contracted according to the geometry of $\Lambda_{d}$.
If the theory is translationally invariant, all $T_n$ are identical, yielding a uniform tensor network.

In general, the tensor elements are determined by the Boltzmann weights, while the tensor indices have a correspondence with the degrees of freedom of the fields.
By contracting the tensors along their shared indices, the nearest-neighbor interaction (or hopping) terms of the original theory are restored.
The dimension of each tensor index is referred to as the bond dimension.
When the theory contains continuous degrees of freedom, the bond dimension is formally infinite. Consequently, the tensor-network representation of $Z$ must be treated as an approximation in practice, which requires the discretization of the degrees of freedom.
It is worth noting that tensors can be constructed not only from the original degrees of freedom of the fields~\cite{Kadoh:2018hqq,Fukuma:2021cni,Nakayama:2024lhb}, but also using dual or auxiliary variables~\cite{Liu:2013nsa,Akiyama:2020sfo,Meurice:2020pxc}.
In LGT, the former construction is convenient when one wishes to impose L\"{u}scher's admissibility condition in a straightforward manner~\cite{Akiyama:2024qer}. 
By contrast, the latter is useful when exploiting global symmetries of the original lattice theory, as it naturally leads to selection rules for the tensor elements~\cite{Meurice:2020gcd}.

The second ingredient is the construction of a coarse-graining transformation $\tau$ from the fundamental tensors to approximately contract the tensor-network representation of $Z$.
Formally, it reads $\tau(\{T^{(l)}_{n}\})=T^{(l+1)}_{n'}$, where $\{T^{(l)}_{n}\}$ denotes the set of local tensors involved in defining the coarse-grained tensor $T^{(l+1)}_{n'}$ at the ($l$+1)-th step, with $T^{(0)}_{n}$ being the initial fundamental tensor.
One of the simplest transformations would be $\tau(T^{(l)}_{n+\hat{\mu}},T^{(l)}_{n})=T^{(l+1)}_{n'}$, where $\tau$ combines two adjacent tensors along the $\mu$-direction into a single coarse-grained tensor $T^{(l+1)}_{n'}$.
Let $D$ be the bond dimension of $T^{(l)}_{n}$.
If the contraction of the two tensors is performed exactly, the resulting $T^{(l+1)}_{n'}$ has bond dimension $D$ along the $\mu$-direction, while in the other directions the bond dimension increases to $D^2$.
Therefore, to make the iterative application of the transformation $\tau$ feasible, one needs to introduce a truncation procedure that reduces the enlarged bond dimension from $D^2$ to a fixed value $\chi$.
The optimal truncation for this purpose is achieved via singular value decomposition (SVD), according to the Eckart--Young theorem~\cite{Eckart_Young_1936}.
\footnote{
This update scheme is known as the higher-order TRG (HOTRG) algorithm~\cite{Xie:2012mjn}.
When the system is translationally invariant on the lattice, this simple update scheme works well in practice.
However, since our goal is to evaluate $Z={\rm tTr}\left[\prod T\right]$, it is more natural to incorporate all tensors $T^{(l)}_{n}$ on the lattice when defining a coarse-grained tensor $T^{(l+1)}_{n'}$.
This amounts to a global optimization of the entire tensor network at a fixed bond dimension $\chi$.
Indeed, several methods have been developed along these lines~\cite{Xie:2009zzd,Morita:2021rmf,Ueda:2025mhu}.
}
We now observe that, after the $l$-time iterative application of the transformation, $2^{l}$ tensors, constituting a $2^{l}$-site system, are approximately contracted.
Since ${\rm tTr}[T^{(l)}_{n}]$ gives the partition function of the corresponding system, the thermodynamic limit can be defined through the convergence of the free energy density with respect to $l$.
This can indeed be regarded as an explicit implementation of the RG concept summarized in Sec.~\ref{sec:Intro}.
It is by reformulating the theory in terms of tensor networks that the SVD can be employed to define a coarse-graining transformation $\tau$.
Moreover, since the SVD provides the optimal rank-$\chi$ approximation and its accuracy improves systematically with increasing $\chi$, the RG procedure is systematically improvable.

\section{Grassmann tensor networks}
\label{sec:GTN}

One of the significant advantages of tensor networks in their application to lattice field theories is that they provide a numerical framework capable of directly handling fermion fields.
Grassmann tensor networks constitute one such framework, where fermions are represented by Grassmann variables, and the TRG method is ready to be applied to their evaluation.
Although the original proposal in Refs.~\cite{Gu:2010yh,Gu:2013gba} introduced Grassmann tensor networks as a way to represent generic quantum states incorporating Grassmann numbers, the formulation can be straightforwardly extended to the Lagrangian formalism.
The first practical application of the Grassmann TRG was to the Schwinger model with Wilson fermions~\cite{Shimizu:2014uva}.
In contrast to conventional MC methods, Grassmann TRG does not have to rely on pseudo-fermion representations, and the Grassmann tensor networks preserve the locality of the original theory. 
Since the local Hilbert space associated with each fermion is finite-dimensional, fermionic systems are particularly well suited to tensor-network representations with finite bond dimension.
For explicit Grassmann tensor-network formulations of relativistic lattice fermions, see Ref.~\cite{Akiyama:2020sfo}.
For a comprehensive review, we refer the reader to Ref.~\cite{Akiyama:2024ush}.

The TRG application to multi-flavor theories constitutes an important step toward their eventual application to QCD. 
Previously, applications of Grassmann tensor networks were restricted to single-flavor theories.
This limitation arises from the exponential growth of the bond dimension with the number of fermion flavors $N_{f}$.
To address this issue, recent proposals treat the flavor index as an additional virtual dimension and construct multilayer Grassmann tensor networks for multi-flavor theories~\cite{Akiyama:2023lvr,Yosprakob:2023tyr}, in which the exponential growth of the bond dimension with $N_{f}$ is absent.

\begin{figure}[t]
    \vspace{-12pt}
    \centering
    \includegraphics[width=0.47\linewidth]{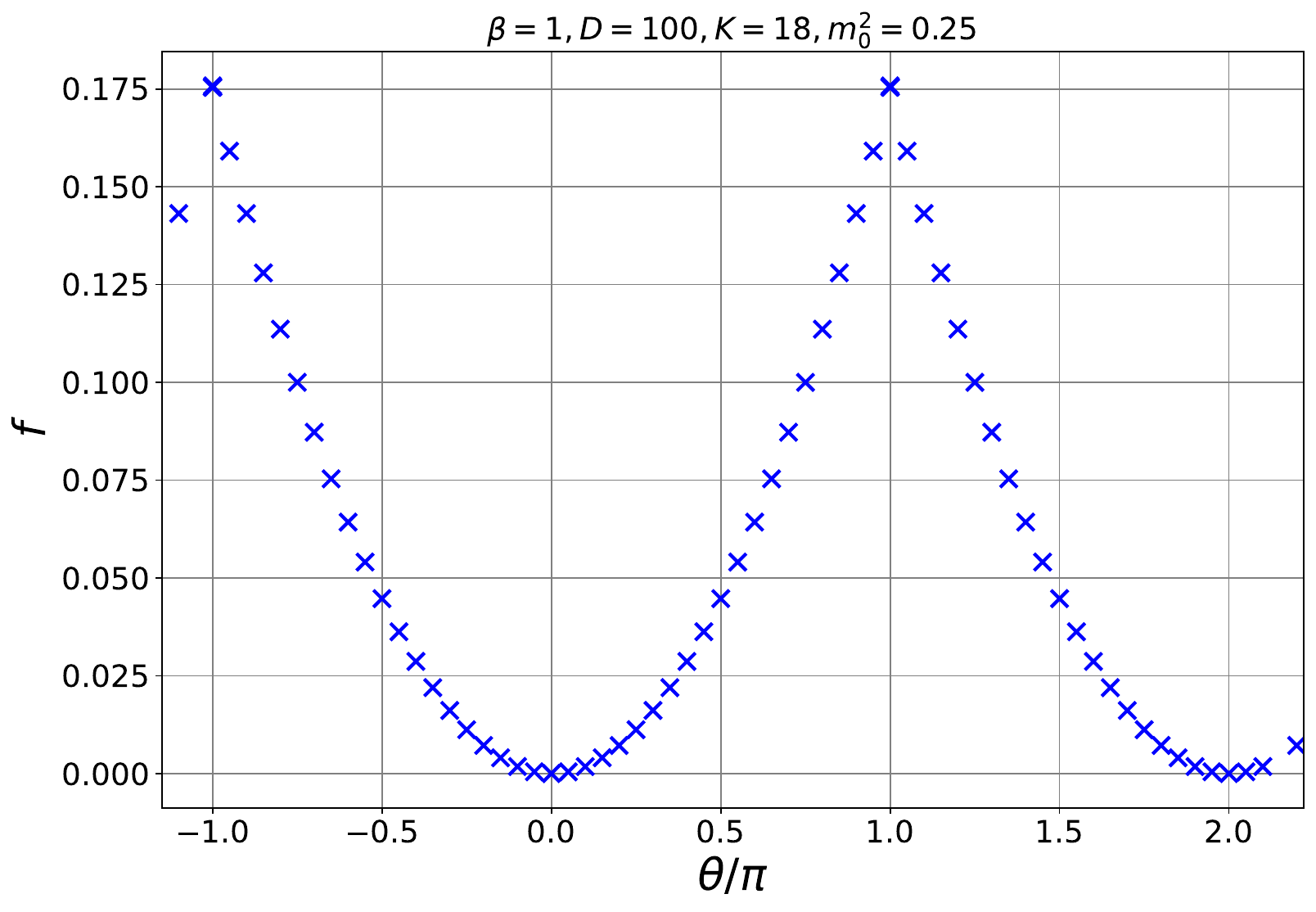}
    \caption{
        Thermodynamic free energy density as a function of $\theta$ for the massive Schwinger model with staggered fermions, adapted from Ref.~\cite{Kanno:2025hgp}.
    }
    \label{fig:schwinger}
\end{figure}

Even in 2D systems, multi-flavor theories provide a variety of physically intriguing problems.
The two-flavor Schwinger model with a $\theta$ term is one such example.
In contrast to the single-flavor case, a recent field-theoretical analysis predicts that the two-flavor Schwinger model exhibits an exponentially small mass gap in the small-mass regime at $\theta=\pi$~\cite{Dempsey:2023gib}.
The Grassmann TRG approach has subsequently been applied to this two-flavor model formulated with staggered fermions~\cite{Kanno:2025hgp}, providing a nonperturbative confirmation of this prediction~\cite{Kanno:2024elz}.
This result crucially relies on the capability of the Grassmann TRG to treat massive fermions. 
In earlier world-line-based TRG~\cite{Butt:2019uul}, non-local sign factors prevented reliable calculations in the massive regime, whereas the Grassmann tensor networks preserve locality and avoid this difficulty, as demonstrated by the $\theta$-dependence of the thermodynamic free energy in the massive regime shown in Fig.~\ref{fig:schwinger}.

\section{Applications to two-color QCD and strong-coupling QCD}

Although the Schwinger model is often regarded as the closest 2D analogue of QCD, it is obviously crucial to extend tensor-network studies to genuinely non-Abelian gauge theories, even in 2D. 
In particular, moving from a $U(1)$ gauge group to $SU(N_{c})$ greatly enlarges the gauge degrees of freedom, and it is far from obvious whether tensor-network methods remain efficient even in 2D non-Abelian gauge theories with dynamical fermions. 

The first TRG study of a non-Abelian gauge theory coupled to fermions is reported in Ref.~\cite{Asaduzzaman:2023pyz}, where $N_{c}=2$ is considered using the reduced staggered fermions~\cite{Catterall:2018pms}.
For the gauge fields, a numerical quadrature method is employed to discretize the path integrals.
The resulting average plaquette as a function of the inverse gauge coupling agrees well with the MC result.
One of the challenges, however, is the rapid growth of the bond dimension: the fundamental tensor carries both gauge and fermionic bond dimensions, $D_{\rm G}$ and $D_{\rm F}$, which combine multiplicatively as $D_{\rm G}\times D_{\rm F}$.
The formulation in Ref.~\cite{Asaduzzaman:2023pyz} leads to $D_{{\rm F}}=2^{N_{\rm hop}\times N_{c}^{2}}$, with $N_{\rm hop}$ the number of hopping terms per spacetime direction.
For reduced staggered fermions, $N_{\rm hop}=1$, which lowers $D_{{\rm F}}$.

Ref.~\cite{Pai:2024tip} proposes strategies to alleviate this difficulty.
First, they introduce an alternative Grassmann tensor-network formulation in which the fermionic bond dimension is reduced to $D_{{\rm F}}=2^{N_{\rm hop}\times N_{c}}$.
Second, they implement a data-compression scheme before the TRG procedure, which substantially lowers the initial bond dimension.
In the strong-coupling regime, the number of tensor elements can typically be reduced to less than $10\%$ of the original size.
These techniques make it possible to investigate two-color QCD with the standard staggered quark action at finite density, including dynamical gauge fields.
The phase structure has been analyzed in terms of the quark number density and chiral and diquark condensates.
Their qualitative behavior in 2D is found to be consistent with mean-field (MF) predictions~\cite{Nishida:2003uj}.
Subsequent work has further demonstrated that the same strategies are efficient for Wilson fermions~\cite{Pai:2025eia}, where a second-order phase transition in the negative-mass region is identified to belong to the 2D Ising universality class.

\begin{figure}[htbp]
    \centering
    \begin{minipage}{0.41\linewidth}
        \centering
        \includegraphics[width=\linewidth]{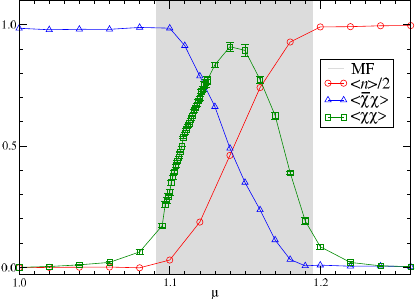}
    \end{minipage}
    \begin{minipage}{0.42\linewidth}
        \centering
        \includegraphics[width=\linewidth]{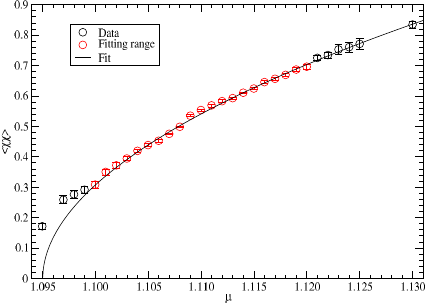}
    \end{minipage}
    \caption{
        TRG results for (3+1)D two-color QCD in the strong-coupling limit, adapted from Ref.~\cite{Sugimoto:2025vui}.
        (Left) Quark number density $\langle n \rangle$, chiral condensate $\langle \bar{\chi}\chi \rangle$, and diquark condensate $\langle \chi\chi \rangle$ as functions of the chemical potential $\mu$. 
        The phase transition points are compared with the MF prediction~\cite{Nishida:2003uj}.
        (Right) The diquark condensate scales as $(\mu-\mu_c)^{\beta_m}$ with $\beta_m=0.514(27)$, consistent with the MF value $\beta_m=1/2$.
    }
    \label{fig:4d_qc2d}
\end{figure}

The formulation constructed in Ref.~\cite{Pai:2024tip} has recently been applied to (3+1)D two-color QCD in the strong-coupling limit~\cite{Sugimoto:2025vui}. 
In this limit, the link variables can be integrated out analytically, and the bond dimension of the resulting tensor network is determined solely by $D_{{\rm F}}$.
In Ref.~\cite{Sugimoto:2025vui}, the phase structure at finite chemical potential and vanishing temperature is investigated using the quark number density, as well as the chiral and diquark condensates. 
The observed critical behavior is found to be consistent with MF theory~\cite{Nishida:2003uj}, as shown in Fig.~\ref{fig:4d_qc2d}.
Furthermore, the approach has been extended to (3+1)D QCD with $N_{c}=3$ at finite density in the strong-coupling limit, marking an important step toward applications to realistic QCD~\cite{Sugimoto:2026wnw}.
The chiral and nuclear transitions are studied with particular focus on the critical endpoint in terms of the quark mass at finite temperature.
The critical mass is determined with reasonable precision, and compared with the MF prediction~\cite{Nishida:2003fb} as well as results from dual simulations~\cite{Kim:2023dnq}.
The study is also extended to the zero-temperature limit, where a clear signal of a first-order phase transition is observed.

Realizing a fully dynamical treatment of gauge fields in (3+1)D constitutes the next important challenge toward tensor-network studies of realistic QCD at finite density and temperature.
Several efforts along this direction have been reported~\cite{Fukuma:2021cni,Kuwahara:2022ubg,Yosprakob:2024sfd,Samberger:2025dmp}.

\section{Partition-function ratios}

Since TRG provides access to the partition function, various thermodynamic quantities can, in principle, be extracted from it.
In addition, Gu and Wen demonstrated that the partition function itself can be useful in certain contexts.
They pointed out that ratios of partition functions can be employed to detect the ground-state degeneracy in systems with discrete global symmetries~\cite{Gu:2009dr}.
One such useful ratio is defined as $X=Z(L_{1},\cdots,L_{d})^{2}/Z(L_{1},\cdots,2L_{d})$,
where $Z$ is the partition function defined on a $d$-dimensional lattice with extent $L_{\mu}$ in the $\mu$-direction, under periodic boundary conditions in all directions.
We refer to $X$ as the Gu--Wen (GW) ratio.
When the global $\mathbb{Z}_{N}$ symmetry is completely spontaneously broken, the GW ratio approaches $N$ in the large-volume limit.
This is because, in that limit, the partition function behaves as $Z\sim (\#~{\rm of~invariant~vacua})\times\exp[-\Lambda L_{1} \cdots L_{d}]$ with a non-universal constant $\Lambda$.
By contrast, $X$ takes the value 1 when the system is in the symmetric phase.
In the large-volume limit, the GW ratio therefore exhibits a discontinuous jump at the transition point.
Consequently, the phase transition associated with the spontaneous breaking of a discrete global symmetry can be identified solely from the GW ratio.

\begin{figure}[htbp]
    \vspace{-10pt}
    \centering
    \includegraphics[width=0.47\linewidth]{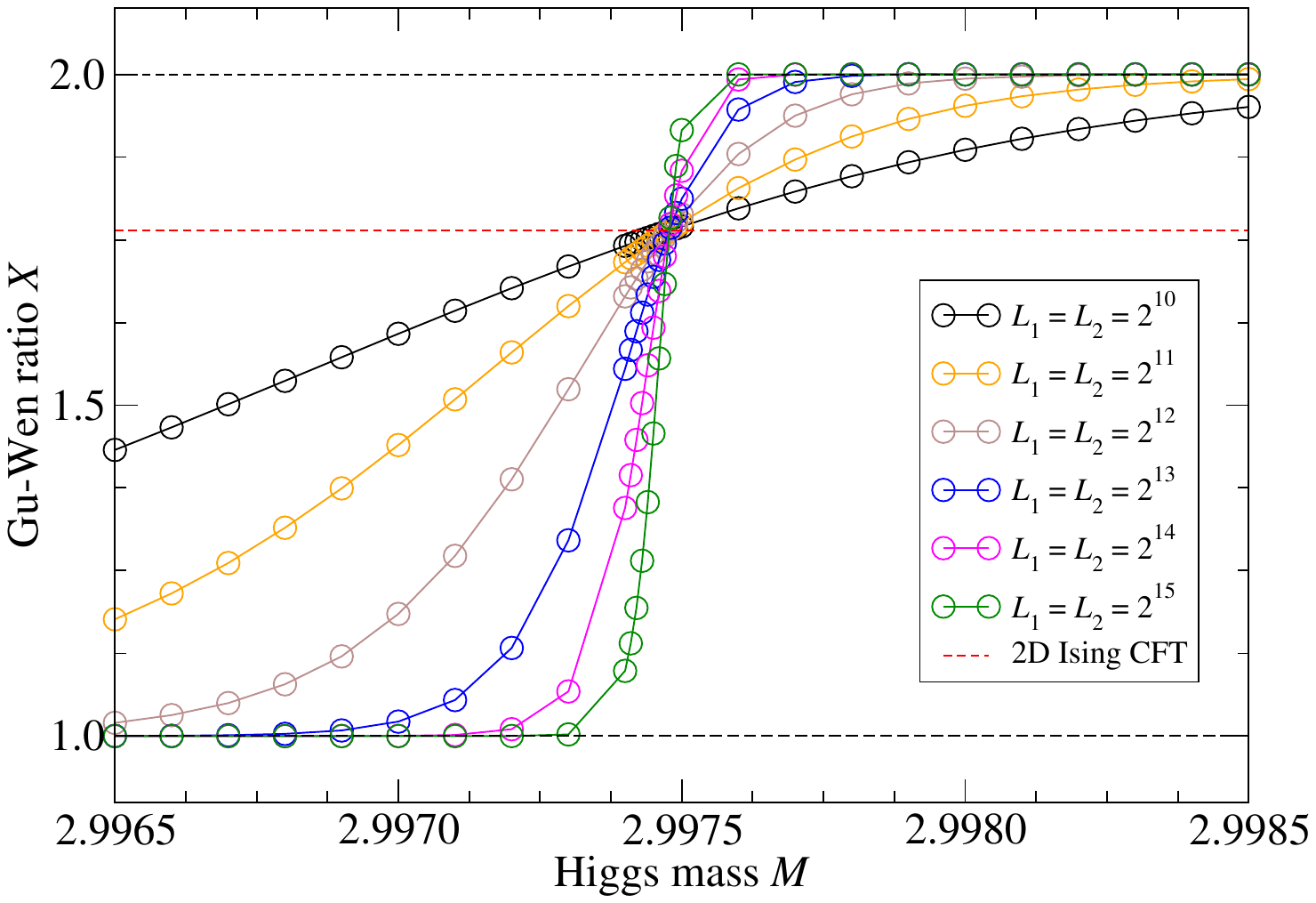}
    \caption{
        GW ratio for the (1+1)D $U(1)$ gauge-Higgs model at $\theta=\pi$ as a function of the Higgs mass $M$.
        The admissibility condition for the $U(1)$ gauge fields is imposed~\cite{Akiyama:2024qer}.
        The phase transition is associated with the spontaneous breaking of $\mathbb{Z}_{2}$ charge conjugation symmetry.
        The value at the scale-invariant point, where data from different volumes intersect, is consistent with the 2D Ising CFT~\cite{Morita:2025hsv}.
    }
    \label{fig:2d_u1}
\end{figure}

Recently, Ref.~\cite{Morita:2024lwg} has reported that the GW ratio obeys the same finite-size scaling formula as the Binder parameter.
Consequently, the GW ratio evaluated on a finite lattice can still be employed to locate the transition point and to extract the shift exponent $1/\nu$.
Subsequent work has argued that the universal value of the GW ratio at criticality is computed from the modular-invariant partition function of the underlying CFT~\cite{Morita:2025hsv}. 
Thus, the GW ratio can be used to identify the universality class of the phase transition.
A concrete example is shown in Fig.~\ref{fig:2d_u1}.
In Ref.~\cite{Aizawa:2025lxi}, the universal value of the GW ratio for the $SU(2)_{k=1}$ Wess--Zumino--Witten CFT has been utilized to investigate the phase diagram of the 2D CP(1) model with a $\theta$ term.

Attempts have also been made to employ ratios of partition functions to study the spontaneous breaking of continuous symmetries.
Ref.~\cite{Akiyama:2024qgv} has argued the application of the GW ratio to cases where a non-Abelian symmetry is spontaneously broken.
Ref.~\cite{Maeda:2025ycr} has proposed that symmetry-twisted partition functions can serve as order parameters for both discrete and continuous global symmetries.
Instead of the GW ratio, they consider the following ratio of partition functions:
$Z_{g}(L_{1},\dots,L_{d})/Z_{1}(L_{1},\dots,L_{d})$, where $Z_{g}$ denotes the partition function with a symmetry twist by an element $g\in G$ of the global symmetry $G$ imposed along the $d$-direction, while periodic boundary conditions are imposed in the other directions.
Note that $Z_{1}$ in the denominator corresponds to the partition function without any twist.
Symmetry-twisted partition functions are then applied to study the $d$-dimensional $O(2)$ model within the TRG framework~\cite{Akiyama:2026dzg}, where it is shown that the phase transition point associated with continuous symmetry breaking can be determined solely from $Z_{g}/Z_{1}$ for $d=3$.
Furthermore, in $d=2$, $Z_g/Z_1$ allows the computation of the helicity modulus, or superfluid density, within the TRG algorithm, entirely without evaluating correlation functions.

We have so far focused on the torus geometry.
In addition, a method to compute partition functions on the Klein bottle and on RP$^2$ has been proposed~\cite{Shimizu:2024ipw}. 
By taking their ratios to the torus partition function, one can extract the crosscap and rainbow free energies, which also encode universal data of the underlying CFT.

\section{CFT data, spectroscopy, and entanglement entropy}
\label{sec:cft}

The transfer matrix is readily obtained by tracing over the spatial indices of the coarse-grained fundamental tensor.
Once the transfer matrix is constructed, CFT data such as the central charge and scaling dimensions can be extracted from its eigenvalue spectrum.
This constitutes one of the key features of the TRG framework, which provides access not only to specific thermodynamic quantities but also directly to universal data.
The CFT data extracted from the transfer matrix provide a practical means of identifying the universality class. 
While well established for studying 2D CFT~\cite{Shimizu:2017onf,Kanno:2024elz,Akiyama:2024qer,Pai:2025eia,Aizawa:2025lxi,Naravane:2026zjk}, this approach has recently been extended to 3D, yielding reasonably accurate scaling dimensions for the 3D Ising model~\cite{Lyu:2024lqh}.

Spectroscopy within the TRG framework has recently progressed substantially.
Refs.~\cite{Ueda:2021vxr,Ueda:2023smj} provide a tensor-network-based level spectroscopy method, which allows for a precise determination of the critical point by exploiting knowledge of the underlying CFT.
Ref.~\cite{Huang:2023vwp} also formulates a tensor-network-based finite-size scaling method that does not require prior knowledge of the underlying CFT.
More recently, Ref.~\cite{Az-zahra:2024gqr} has enabled the identification of both the quantum numbers and the energy spectrum, from which scattering phase shifts were extracted using L\"{u}scher's formula.

The transfer-matrix formalism can be further utilized to construct a path-integral representation of the thermal density matrix, which allows for the computation of thermal and entanglement entropies~\cite{Yang:2015rra} and the R\'{e}nyi entropy~\cite{Bazavov:2017hzi}.
More recently, a construction of the density matrix for the (1+1)D $O(3)$ nonlinear sigma model is reported~\cite{Luo:2023ont}.
The central charge $c$ is extracted from the asymptotic scaling of the entropies toward the continuum limit and is found to be consistent with $c=2$.
Ref.~\cite{Hayazaki:2025srr} has further developed a method to compute the entanglement entropy for subsystems of arbitrary size.

\section{Synergistic and hybrid computational approaches}
\label{sec:synergy}

Recently, the TRG approach has also been extended through its integration with other numerical algorithms, leading to various hybrid frameworks.

One such example is the combination of TRG with stochastic methods, originally proposed in Ref.~\cite{ferris2015unbiasedmontecarloage}. 
The key idea is to replace the truncated SVD, which keeps only the $\chi$ largest singular values, with a randomized truncation scheme, thereby trading the systematic error from finite bond dimension $\chi$ for a controllable statistical error. 
This approach was further developed in Ref.~\cite{Huggins:2017yzz}, demonstrating that, despite reduced accuracy in individual samples, it can yield unbiased error estimates and introduce an additional control parameter beyond $\chi$.
Ref.~\cite{Arai:2022uee} has proposed a common-noise method whose computational cost scales logarithmically with the system volume. 
Although it introduces a systematic error due to noise correlations, a simple functional form in terms of the number of noise samples can work for enabling controlled error reduction in practice.
Furthermore, Ref.~\cite{Todo:2024fzs} has formulated a Markov chain MC method directly on the TRG-generated tensor network, alleviating the sign problem and improving the average sign as $\chi$ increases.

Automatic differentiation (AD) is another important extension incorporated into tensor-network algorithms, including TRG methods~\cite{Liao:2019bye,Chen:2019mip}. Since AD evaluates derivatives with machine precision, thermodynamic quantities can be computed without additional approximations beyond the original coarse-graining at fixed $\chi$. 
While reverse-mode AD has commonly been employed, Ref.~\cite{Sugimoto:2026zxv} has pointed out that forward-mode AD is more advantageous for incorporation into TRG.

Since high accuracy in TRG requires a large bond dimension $\chi$, and the computational cost scales as a power of $\chi$ with an exponent that grows rapidly with the spacetime dimension, efficient use of HPCI is indispensable, particularly for future QCD applications. 
MPI parallelization on CPUs has already been developed~\cite{Akiyama:2019chk,Yamashita:2021yxs}, and GPU acceleration~\cite{Jha:2023bpn}, including multi-GPU implementations~\cite{Sugimoto:2025xva}, provides a promising path toward large-scale computations.

Finally, TRG has been explored for simulating real-time dynamics. 
The first attempt was made in 2019~\cite{Takeda:2019idb}, followed by subsequent work computing real-time correlators in the (1+1)D $\phi^4$ theory~\cite{Takeda:2021mnc}. 
More recently, Ref.~\cite{Hite:2024ulb} has developed a TRG-based algorithm for real-time evolution of quantum many-body systems. 
These developments will open a new avenue toward bridging Lagrangian- and Hamiltonian-based approaches, with potential connections to quantum computing in the future.

\section{Summary and outlook}
\label{sec:summary}

Demonstrating the TRG approach as a concrete realization of the RG concept established in the literature, we have reviewed several recent advances in this field.
Grassmann tensor networks offer a suitable framework for applying TRG to non-Abelian gauge theories with dynamical fermions, with applications to two-color QCD and strong-coupling QCD reported in both (1+1)D and (3+1)D. 
Since TRG directly computes partition functions, their ratios provide a practical tool for investigating phase transitions. 
In contrast to conventional MC methods, TRG enables direct access to CFT data via the transfer-matrix formalism. 
Recent developments also include spectroscopy methods and computational strategies for entanglement entropy, as well as synergies with other algorithms, such as stochastic techniques and AD, which broaden the scope of TRG applications.

These developments reflect active cross-disciplinary interactions among high-energy physics, condensed-matter physics, and computational science. 
In view of future applications to QCD, further methodological developments are still required, including improvements in numerical stability, scalability, and the treatment of gauge degrees of freedom.
Advances driven by applications to lattice QCD would also influence broader studies of quantum many-body systems. 
Continued progress is therefore expected through sustained collaboration across disciplines and research communities.

\begin{acknowledgments}
    S. A. acknowledges the support from JSPS KAKENHI (JP23K13096 and JP25H01510), the Center of Innovations for Sustainable Quantum AI (JST Grant Number JPMJPF2221), the Endowed Project for Quantum Software Research and Education, the University of Tokyo, and the Top Runners in Strategy of Transborder Advanced Researches (TRiSTAR) program conducted as the Strategic Professional Development Program for Young Researchers by the MEXT. 
\end{acknowledgments}

\bibliographystyle{JHEP}
\bibliography{bib/ref}

\end{document}